\documentclass[3p,times]{elsarticle}

\usepackage{ecrc}

\usepackage{epstopdf}

\volume{00}

\firstpage{1}

\journalname{Nuclear Physics A}

\runauth{Amaresh Jaiswal}

\jid{nupha}

\jnltitlelogo{Nuclear Physics A}

\usepackage{graphicx}
\usepackage{amsmath,amssymb}

\begin{document}

\begin{frontmatter}



\title{Relaxation-time approximation and relativistic third-order viscous hydrodynamics from kinetic theory}

\author{Amaresh Jaiswal}
\address{Tata Institute of Fundamental Research, Mumbai 400005, India}




\begin{abstract}
Using the iterative solution of Boltzmann equation in the 
relaxation-time approximation, the derivation of a third-order 
evolution equation for shear stress tensor is presented. To this end 
we first derive the expression for viscous corrections to the 
phase-space distribution function, $f(x,p)$, up to second-order in 
derivative expansion. The expression for $\delta f(x,p)$ obtained in 
this method does not lead to violation of the experimentally 
observed $1/\sqrt{m_T}$ scaling of the femtoscopic radii, as opposed 
to the widely used Grad's 14-moment approximation. Subsequently, we 
present the derivation of a third-order viscous evolution equation 
and demonstrate the significance of this derivation within 
one-dimensional scaling expansion. We show that results obtained 
using third-order evolution equations are in excellent accordance 
with the exact solution of Boltzmann equation as well as with 
transport results.
\end{abstract}


\end{frontmatter}




\section{Introduction}

Hydrodynamics is an effective theory used to describe the 
long-wavelength, low-frequency limit of the microscopic dynamics of 
a many-body system, which is close to equilibrium. Relativistic 
hydrodynamics is formulated as derivative expansion where ideal 
hydrodynamics is of zeroth-order. The first-order dissipative 
theories (collectively known as relativistic Navier-Stokes theory 
\cite{Eckart:1940zz,Landau}) suffer from acausality and numerical 
instability. Although the acausality problem is rectified in the 
second-order Israel-Stewart (IS) theory \cite{Israel:1979wp}, 
stability is not guaranteed. The IS hydrodynamics has been used 
quite extensively to study the collective behaviour of hot and 
dense, strongly interacting matter created in high-energy heavy-ion 
collision experiments at the Relativistic Heavy-Ion Collider (RHIC) 
and at the Large Hadron Collider (LHC).

Despite its success in explaining a wide range of collective 
phenomena observed in ultra-relativistic heavy-ion collisions, the 
formulation of IS theory is based on several approximations and 
assumptions. Israel and Stewart assumed Grad's 14-moment 
approximation for the non-equilibrium distribution function and 
obtained the dissipative evolution equations from the second moment 
of the Boltzmann equation \cite{Israel:1979wp}. It was shown 
recently that iterative solution of the Boltzmann equation can be 
used instead of 14-moment approximation \cite{Jaiswal:2013npa} and 
the dissipative equations can be derived directly from their 
definitions without resorting to an arbitrary choice of moment of 
the Boltzmann equation \cite{Denicol:2010xn}. Apart from these 
problems in the theoretical formulation, IS theory suffers from 
several other shortcomings. In one-dimensional scaling expansion 
\cite{Bjorken:1982qr}, IS theory results in unphysical effects such 
as negative longitudinal pressure \cite{Martinez:2009mf} and 
reheating of the expanding medium \cite{Muronga:2003ta}. Moreover, 
comparison of results obtained using IS evolution equation with 
transport results show disagreement for $\eta/s>0.5$ indicating the 
breakdown of second-order theory \cite{El:2008yy}. Furthermore, 
inclusion of viscous corrections to the phase-space distribution 
function, $f(x,p)$, via 14-moment approximation results in the 
violation of experimentally observed and ideal hydrodynamic 
prediction of $1/\sqrt{m_T}$ scaling of the longitudinal Hanbury 
Brown-Twiss (HBT) radii \cite{Teaney:2003kp}.

In order to widen the range of applicability of the IS theory, 
second-order dissipative equations was derived from the Boltzmann 
equation where the collision term was generalized to incorporate 
nonlocal effects through gradients of $f(x,p)$ \cite 
{Jaiswal:2012qm}. Moreover, it was also shown that inclusion of 
third-order corrections to shear evolution equation led to an 
improved agreement with the transport results \cite 
{El:2009vj,Jaiswal:2013vta}. Furthermore, to improve Grad's 
14-moment approximation beyond its current scope, a general moment 
method was devised by introducing orthogonal basis in momentum 
series expansion \cite{Denicol:2012cn}. The accurate and consistent 
formulation of the theory of relativistic viscous hydrodynamics is 
not yet conclusively resolved and is presently a topic of intense 
investigation \cite{Jaiswal:2013npa,Denicol:2010xn,Jaiswal:2012qm, 
El:2009vj,Jaiswal:2013vta,Denicol:2012cn,Jaiswal:2013fc, 
Bhalerao:2013aha,Bhalerao:2013pza}.


\section{Relativistic hydrodynamics}

The equation of motion governing the hydrodynamic evolution of a 
relativistic system with no net conserved charges is obtained from 
the local conservation of energy and momentum, $\partial_\mu 
T^{\mu\nu}=0$. In terms of single-particle phase-space distribution 
function, the energy-momentum tensor of a macroscopic system can be 
expressed as \cite{deGroot}
\begin{align}\label{NTD}
T^{\mu\nu} &= \!\int\! dp \ p^\mu p^\nu\, f(x,p) = \epsilon u^\mu u^\nu 
- P\Delta ^{\mu \nu} + \pi^{\mu\nu},
\end{align}
where $dp\equiv g d{\bf p}/[(2 \pi)^3|\bf p|]$, $g$ being the 
degeneracy factor, $p^\mu$ is the particle four-momentum, and 
$f(x,p)$ is the phase-space distribution function. In the tensor 
decomposition, $\epsilon$, $P$, and $\pi^{\mu\nu}$ are energy 
density, thermodynamic pressure, and shear stress tensor, 
respectively. The projection operator $\Delta^{\mu\nu}\equiv 
g^{\mu\nu}-u^\mu u^\nu$ is orthogonal to the hydrodynamic 
four-velocity $u^\mu$ defined in the Landau frame: $T^{\mu\nu} 
u_\nu=\epsilon u^\mu$. The metric tensor is Minkowskian, 
$g^{\mu\nu}\equiv\mathrm{diag}(+,-,-,-)$. Here we restrict 
ourselves to a system of massless particles (ultrarelativistic 
limit) for which the bulk viscosity vanishes.

The conservation of the energy-momentum tensor, when projected along 
and orthogonal to $u^\mu$, leads to the evolution equations for 
$\epsilon$ and $u^\mu$:
\begin{align}\label{evol}
\dot\epsilon + (\epsilon+P)\theta - \pi^{\mu\nu}\sigma_{\mu\nu} &= 0, \qquad
(\epsilon+P)\dot u^\alpha - \nabla^\alpha P + \Delta^\alpha_\nu \partial_\mu \pi^{\mu\nu}  = 0,
\end{align}
where we employ the standard notation $\dot A\equiv 
u^\mu\partial_\mu A$ for comoving derivative, $\theta\equiv 
\partial_\mu u^\mu$ for expansion scalar, 
$\sigma^{\mu\nu}\equiv(\nabla^\mu u^\nu+\nabla^\nu 
u^\mu)/2-(\theta/3)\Delta^{\mu\nu}$ for velocity stress tensor, and 
$\nabla^\alpha\equiv\Delta^{\mu\alpha}\partial_\mu$ for space-like 
derivatives. For the massless case, the equation of state relating 
energy density and pressure is $\epsilon=3P\propto\beta^{-4}$. The 
matching condition $\epsilon=\epsilon_0$ is employed to fix the 
inverse temperature, $\beta\equiv1/T$, where $\epsilon_0$ is the 
equilibrium energy density. The derivatives of $\beta$, 
\begin{align}\label{evolb}
\dot\beta &= \frac{\beta}{3}\theta - \frac{\beta}{12P}\pi^{\rho\gamma}\sigma_{\rho\gamma}, \qquad
\nabla^\alpha\beta = -\beta\dot u^\alpha - \frac{\beta}{4P} \Delta^\alpha_\rho \partial_\gamma \pi^{\rho\gamma}, 
\end{align}
can be obtained from Eq. (\ref{evol}). When the system is close to 
local thermodynamic equilibrium, the distribution function can be 
written as $f=f_0+\delta f$, where $\delta f\ll f_0$, 
$f_0=\exp(-\beta\,u\cdot p)$ is the equilibrium distribution 
function of Boltzmann particles at vanishing chemical potential and 
$u\cdot p\equiv u_\mu p^\mu$. Projecting the traceless symmetric 
part of Eq. (\ref{NTD}) using the operator 
$\Delta^{\mu\nu}_{\alpha\beta}\equiv 
(\Delta^{\mu}_{\alpha}\Delta^{\nu}_{\beta} + 
\Delta^{\mu}_{\beta}\Delta^{\nu}_{\alpha})/2 - 
(1/3)\Delta^{\mu\nu}\Delta_{\alpha\beta}$, we can write the shear 
stress tensor and its time evolution as,
\begin{align}\label{FSE}
\pi^{\mu\nu} &= \Delta^{\mu\nu}_{\alpha\beta} \int dp \, p^\alpha p^\beta\, \delta f, \qquad
\dot\pi^{\langle\mu\nu\rangle} = \Delta^{\mu\nu}_{\alpha\beta} \int dp\, p^\alpha p^\beta\, \delta\dot f.
\end{align}
In the following we obtain $\delta f$ and derive evolution equation 
for shear stress tensor in terms of hydrodynamic variables.


\section{Viscous evolution equations}

We start from the relativistic Boltzmann equation with 
relaxation-time approximation for the collision term \cite 
{Anderson_Witting},
\begin{equation}\label{RBE}
p^\mu\partial_\mu f = -\left(u\!\cdot\! p\right) \frac{\delta f}{\tau_R} \quad\Rightarrow\quad
f=f_0-\frac{\tau_R}{(u\!\cdot\! p)}\,p^\mu\partial_\mu f,
\end{equation}
where $\tau_R$ is the relaxation time. Expanding the distribution 
function $f$ about its equilibrium value in powers of space-time 
gradients, i.e., $f = f_0 + \delta f^{(1)} + \delta f^{(2)} + \cdots$
and solving Eq. (\ref {RBE}) iteratively, we obtain \cite
{Jaiswal:2013npa,Jaiswal:2013vta},
\begin{align}
\delta f^{(1)} &= -\frac{\tau_R}{u\cdot p} \, p^\mu \partial_\mu f_0, \qquad
\delta f^{(2)} = \frac{\tau_R}{u\cdot p}p^\mu p^\nu\partial_\mu\left(\frac{\tau_R}{u\cdot p} \partial_\nu f_0\right). \label{SOC}
\end{align}
Substituting $\delta f=\delta f^{(1)}$ in the expression for 
$\pi^{\mu\nu}$ in Eq. (\ref{evolb}), performing the integrations, 
and retaining only first-order terms, we obtain $\pi^{\mu\nu} = 
2\tau_R\beta_\pi\sigma^{\mu\nu}$, where $\beta_\pi = 4P/5$ 
\cite{Jaiswal:2013npa}.

To obtain the second-order evolution equation for shear stress 
tensor, we rewrite Eq. (\ref{RBE}) in the form $\delta\dot f = -\dot 
f_0 - p^\gamma\nabla_\gamma f/(u\!\cdot\! p) - \delta f/\tau_R$. 
Using this expression for $\delta\dot f$ in Eq. (\ref{FSE}), we obtain
\begin{equation} 
\dot\pi^{\langle\mu\nu\rangle} + \frac{\pi^{\mu\nu}}{\tau_R} = - \Delta^{\mu\nu}_{\alpha\beta} 
\int dp \, p^\alpha p^\beta \left(\dot f_0 + \frac{1}{u\!\cdot\! p}p^\gamma\nabla_\gamma f\right). \label{SOSE}
\end{equation}
Using Eq. (\ref{SOC}) for $\delta f^{(1)}$ and Eq. (\ref{evolb}) 
for derivatives of $\beta$, and keeping terms up to quadratic order 
in gradients, the second-order shear evolution equation is obtained 
as \cite {Jaiswal:2013npa}
\begin{equation}\label{SOSHEAR}
\dot{\pi}^{\langle\mu\nu\rangle} + \frac{\pi^{\mu\nu}}{\tau_\pi} = 
2\beta_{\pi}\sigma^{\mu\nu}
+2\pi_\gamma^{\langle\mu}\omega^{\nu\rangle\gamma}
-\frac{10}{7}\pi_\gamma^{\langle\mu}\sigma^{\nu\rangle\gamma} 
-\frac{4}{3}\pi^{\mu\nu}\theta,
\end{equation}
where $\omega^{\mu\nu}\equiv(\nabla^\mu u^\nu-\nabla^\nu u^\mu)/2$ 
is the vorticity tensor. Using Eqs. (\ref {evolb}) and (\ref 
{SOSHEAR}) in Eq. (\ref {SOC}), we arrive at \cite{Bhalerao:2013pza},
\begin{align}\label{SOVC}
\delta f =\ &  \frac{f_0\beta}{2\beta_\pi(u\!\cdot\!p)} p^\alpha\! p^\beta \pi_{\alpha\beta}
-\frac{f_0\beta}{\beta_\pi} \bigg[\frac{\tau_\pi}{u\!\cdot\!p} p^\alpha p^\beta \pi^\gamma_\alpha\omega_{\beta\gamma} 
+\frac{\tau_\pi}{3(u\!\cdot\!p)} p^\alpha p^\beta \pi_{\alpha\beta}\theta
+\frac{(u\!\cdot\!p)}{70\beta_\pi} \pi^{\alpha\beta}\pi_{\alpha\beta} 
-\frac{6\tau_\pi}{5} p^\alpha\dot u^\beta\pi_{\alpha\beta} 
+\frac{\tau_\pi}{5} p^\alpha \!\left(\nabla^\beta\pi_{\alpha\beta}\right) \\
&-\frac{5}{14\beta_\pi (u\!\cdot\!p)} p^\alpha p^\beta \pi^\gamma_\alpha\, \pi_{\beta\gamma}   
-\frac{3\tau_\pi}{(u\!\cdot\!p)^2} p^\alpha p^\beta p^\gamma \pi_{\alpha\beta}\dot u_\gamma 
+\frac{\tau_\pi}{2(u\!\cdot\!p)^2} p^\alpha p^\beta p^\gamma\! \left(\nabla_\gamma\pi_{\alpha\beta}\right) 
-\frac{\beta+(u\!\cdot\!p)^{-1}}{4(u\!\cdot\!p)^2\beta_\pi}\! \left(p^\alpha p^\beta \pi_{\alpha\beta}\right)^2\bigg]
+{\cal O}(\delta^3), \nonumber
\end{align}
where the first term on the right-hand side of the above equation 
corresponds to first-order correction whereas the terms within 
square brackets are of second order. It is straightforward to show 
that the form of $\delta f$ in Eq. (\ref {SOVC}) is consistent with 
the definition of the shear stress tensor, Eq. (\ref{FSE}), and 
satisfies the matching condition $\epsilon =\epsilon_0$ and the 
Landau frame definition $u_\nu T^{\mu \nu} = \epsilon u^\mu$ at each 
order \cite{Bhalerao:2013pza}. It is important to note that the 
experimentally observed $1/\sqrt{m_T}$ scaling of the longitudinal 
HBT radii \cite{Bearden:2001sy}, also predicted by ideal 
hydrodynamics, is violated by incorporating viscous correction 
through Grad's 14-moment approximation \cite{Teaney:2003kp}. However 
the form of $\delta f$ given in Eq. (\ref {SOVC}) does not lead to 
such undesirable effects \cite{Bhalerao:2013pza}.

To obtain a third-order evolution equation for shear stress tensor, 
we substitute $\delta f$ from Eq. (\ref{SOVC}) in Eq. (\ref{SOSE}). 
Keeping terms up to cubic order in derivatives, after a 
straightforward but tedious algebra, we finally obtain a third-order 
evolution equation for shear stress tensor \cite{Jaiswal:2013vta}: 
\begin{align}\label{TOSHEAR}
\dot{\pi}^{\langle\mu\nu\rangle} \!=& -\frac{\pi^{\mu\nu}}{\tau_\pi}
+2\beta_\pi\sigma^{\mu\nu}\!
+2\pi_{\gamma}^{\langle\mu}\omega^{\nu\rangle\gamma}\!
-\frac{10}{7}\pi_\gamma^{\langle\mu}\sigma^{\nu\rangle\gamma}\! 
-\frac{4}{3}\pi^{\mu\nu}\theta
+\frac{25}{7\beta_\pi}\pi^{\rho\langle\mu}\omega^{\nu\rangle\gamma}\pi_{\rho\gamma}\!
-\frac{1}{3\beta_\pi}\pi_\gamma^{\langle\mu}\pi^{\nu\rangle\gamma}\theta 
-\frac{38}{245\beta_\pi}\pi^{\mu\nu}\pi^{\rho\gamma}\sigma_{\rho\gamma} \nonumber \\
&-\frac{22}{49\beta_\pi}\pi^{\rho\langle\mu}\pi^{\nu\rangle\gamma}\sigma_{\rho\gamma} 
-\frac{24}{35}\nabla^{\langle\mu}\left(\pi^{\nu\rangle\gamma}\dot u_\gamma\tau_\pi\right)
+\frac{4}{35}\nabla^{\langle\mu}\left(\tau_\pi\nabla_\gamma\pi^{\nu\rangle\gamma}\right) 
-\frac{2}{7}\nabla_{\gamma}\left(\tau_\pi\nabla^{\langle\mu}\pi^{\nu\rangle\gamma}\right)
+\frac{12}{7}\nabla_{\gamma}\left(\tau_\pi\dot u^{\langle\mu}\pi^{\nu\rangle\gamma}\right) \\
&-\frac{1}{7}\nabla_{\gamma}\left(\tau_\pi\nabla^{\gamma}\pi^{\langle\mu\nu\rangle}\right)
+\frac{6}{7}\nabla_{\gamma}\left(\tau_\pi\dot u^{\gamma}\pi^{\langle\mu\nu\rangle}\right)
-\frac{2}{7}\tau_\pi\omega^{\rho\langle\mu}\omega^{\nu\rangle\gamma}\pi_{\rho\gamma}
-\frac{2}{7}\tau_\pi\pi^{\rho\langle\mu}\omega^{\nu\rangle\gamma}\omega_{\rho\gamma} 
-\frac{10}{63}\tau_\pi\pi^{\mu\nu}\theta^2
+\frac{26}{21}\tau_\pi\pi_\gamma^{\langle\mu}\omega^{\nu\rangle\gamma}\theta. \nonumber
\end{align}
This is the main result of the present work. We compare the above 
equation with that obtained in Ref. \cite {El:2009vj} by invoking 
the second law of thermodynamics,
\begin{align}\label{TOEF}
\dot\pi^{\langle\mu\nu\rangle} =& -\frac{\pi^{\mu\nu}}{\tau_\pi'} + 2\beta_\pi'\sigma^{\mu\nu}
-\frac{4}{3}\pi^{\mu\nu}\theta 
+ \frac{5}{36\beta_\pi'}\pi^{\mu\nu}\pi^{\rho\gamma}\sigma_{\rho\gamma}
-\frac{16}{9\beta_\pi'}\pi^{\langle\mu}_{\gamma}\pi^{\nu\rangle\gamma}\theta,
\end{align}
where $\beta_\pi'=2P/3$ and $\tau_\pi'=\eta/\beta_\pi'$. We notice 
that the right-hand-side of the above equation contains one 
second-order and two third-order terms compared to three 
second-order and fourteen third-order terms obtained in Eq. (\ref 
{TOSHEAR}). This confirms the fact that the evolution equation 
obtained by invoking the second law of thermodynamics is incomplete.


\section{Numerical results and conclusions}

In the following, we consider boost-invariant Bjorken expansion of a 
massless Boltzmann gas \cite{Bjorken:1982qr}. We have solved the 
evolution equations with initial temperature $T_0=300$ MeV at 
initial time $\tau_0=0.25$ fm/c and with $T_0=500$ MeV at 
$\tau_0=0.4$ fm/c, corresponding to initial conditions of RHIC and 
LHC, respectively.

\begin{figure}[t]
 \begin{center}
  \scalebox{.35}{\includegraphics{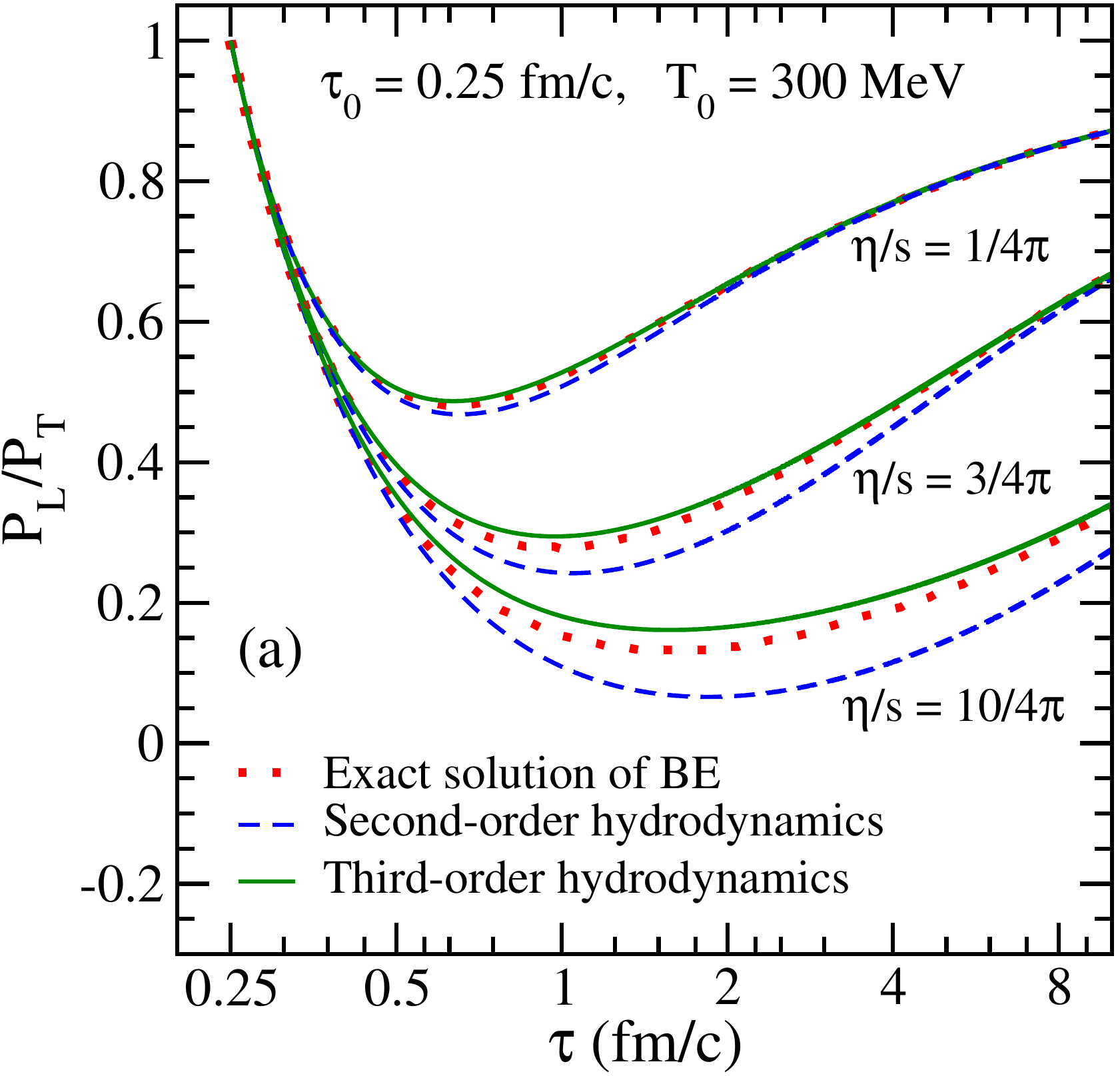}}\hfil
  \scalebox{.35}{\includegraphics{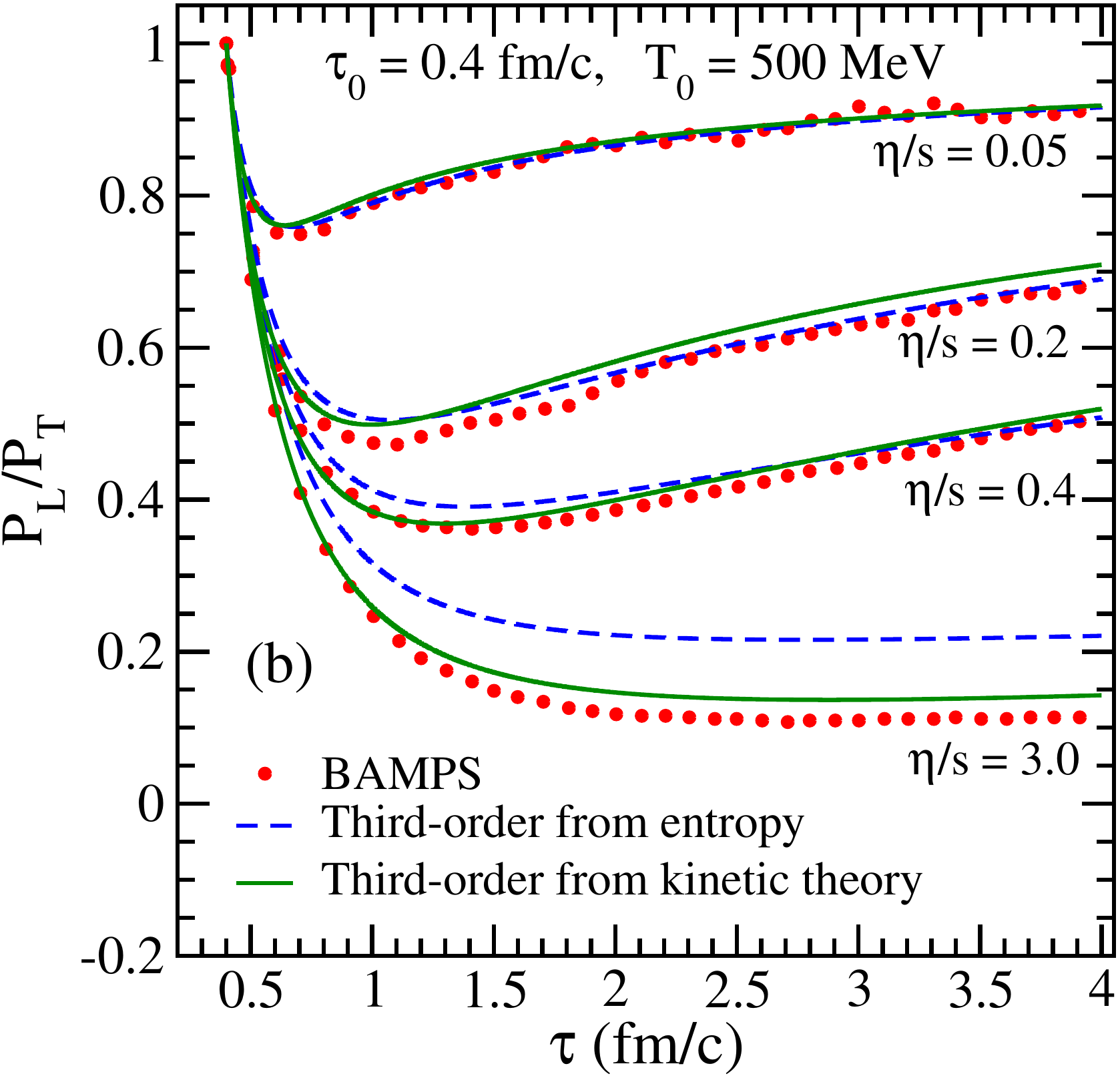}}
 \end{center}
 \vspace{-0.6cm}
 \caption{(a): Time evolution of $P_L/P_T$ obtained 
  using exact solution of Boltzmann equation (dotted line), 
  second-order equations (dashed lines), and third-order equations 
  (solid lines). (b): Time evolution of $P_L/P_T$ in BAMPS (dots), 
  third-order calculation from entropy method, Eq. (\ref{TOEF}) 
  (dashed lines), and the present work (solid lines). Both figures 
  are for isotropic initial pressure configuration $(\pi_0=0)$ and 
  various $\eta/s$.}
 \label{PLPT}
\end{figure}

Figures \ref{PLPT} (a) and (b) shows the proper time dependence of 
pressure anisotropy $P_L/P_T\equiv(P-\pi)/(P+\pi/2)$ where 
$\pi\equiv-\tau^2\pi^{\eta\eta}$. In Fig. \ref {PLPT} (a), we 
observe an improved agreement of third-order results (solid lines) 
with the exact solution of Boltzmann equation (dotted line) \cite 
{Florkowski:2013lza} as compared to second-order results (dashed 
line) suggesting the convergence of the derivative expansion. In 
Fig. \ref{PLPT} (b) we notice that while the results from entropy 
derivation (dashed lines) overestimate the pressure anisotropy for 
$\eta/s>0.2$, those obtained in the present work (solid lines) are 
in better agreement with the results of the parton cascade BAMPS 
(dots) \cite{El:2009vj,Xu:2004mz}.

To summarize, we have derived a third-order evolution equation for 
the shear stress tensor from kinetic theory. We iteratively solved 
the Boltzmann equation in relaxation time approximation to obtain 
the non-equilibrium distribution function up to second-order in 
gradients. Using this form of the non-equilibrium distribution 
function, instead of Grad's 14-moment approximation, the evolution 
equation for shear tensor was derived directly from its definition. 
Within one-dimensional scaling expansion we demonstrated that the 
third-order viscous hydrodynamic equation derived here provides a 
very good approximation to the exact solution of Boltzmann equation. 
We also showed that our results are in better agreement with BAMPS 
compared to third-order viscous hydrodynamics derived from entropy.








\end{document}